\documentclass[twocolumn,showpacs,preprintnumbers,amsmath,amssymb,pre]{revtex4-1}

\usepackage{color}    
\usepackage{graphicx}
\usepackage{dcolumn}
\usepackage{bm}
\usepackage{subfigure}
\usepackage{amssymb}
\usepackage{multirow}

\graphicspath{{plots/}}

\begin{document}


\title{A critique of ``quantum dusty plasmas''}

\author{Zh.~A. Moldabekov$^{1,2,3}$,~M. Bonitz$^1$,~T.~S. Ramazanov$^2$}%
\affiliation{
 $^1$Institut f\"ur Theoretische Physik und Astrophysik, Christian-Albrechts-Universit\"at zu Kiel,
 Leibnizstra{\ss}e 15, 24098 Kiel, Germany}
 \affiliation{
 $^2$Institute for Experimental and Theoretical Physics, Al-Farabi Kazakh National University, 71 Al-Farabi str.,  
  050040 Almaty, Kazakhstan
  }
\affiliation{$^3$Institute of Applied Sciences and IT, 40-48 Shashkin Str., 050038 Almaty, Kazakhstan}


\begin{abstract}
During the recent decade a steadily increasing number of papers devoted to 
so-called ``quantum dusty plasmas'' (QDP) appeared. These systems are combining 
the properties of dusty (i.e. micrometer to nanometer-sized particles 
containing) plasmas  with quantum effects 
appearing at 
high-density and/or low-temperature. Many exciting properties of QDP were 
predicted including 
nontrivial collective oscillations and linear and nonlinear 
waves. It was predicted that the results have relevance for dense astrophysical 
plasmas such as white dwarf stars, neutron stars, magnetars and laboratory plasmas in micro- and 
nano-devices. Here we present a careful examination of the plasma parameters 
for the expected applications and compare them to the conditions under which a 
dust particle can exist in a quantum plasma. The conclusion is that quantum dusty 
plasmas do not exist in reality. 

\end{abstract}

\maketitle

 \section{Introduction}\label{s:intro}
 Dusty plasmas have been an active research field for more than 25 years. The 
addition of micrometer-size ``dust'' particles to a low-temperature plasma 
leads to a broad range of novel phenomena, e.g.~\cite{loewen-morfill}. They 
include strong correlation effects, due to the high charge (on the order of 
several thousand elementary charges) of these particles, 
e.g.~\cite{bonitz_rpp10}, and interesting transport and wave phenomena 
\cite{shukla-dust-book}.
 A second active direction of modern plasma physics is quantum plasmas which 
are observed at low temperature or if matter is very highly compressed, for the 
relevant parameter range, see Fig.~\ref{fig:1}. For example, dense plasmas 
with temperatures comparable to the electronic Fermi 
energy (see below) are produced by compression of  matter up to (or 
even beyond) the densities corresponding to pressure ionization of atoms, e.g. 
\cite{graziani-book}. The behavior of these systems is governed by electronic 
quantum effects, so their accurate description is of high importance and constitutes an actively developing field, see Ref. 
\cite{dornheim_physrep18}, for a recent overview. The investigation of dense 
plasmas (or ``warm dense matter'') is motivated by their relevance 
for understanding the physics of  massive 
astrophysical objects such as giant planets, white dwarf or neutron stars, 
but also by laboratory applications such as laser plasma interaction and  
inertial confinement fusion. Examples of the relevant 
experimental plasma parameters can be found, e.g., 
in Refs.~\cite{Fortov2016, PRE2018}.
 
 It is, therefore, an interesting idea to combine dusty plasmas and quantum plasmas by ``adding'' dust particles to a quantum plasma or by cooling a dusty plasma to low temperatures, giving rise to a new field of ``quantum dusty plasma'' (QDP). As in the case of dusty plasmas, this promises many novel physical phenomena. 
In fact, this idea was put forward in 2005 by Shukla and Ali  \cite{Ali2005}. This paper has had a high impact in the field (it collected more than 90 and the follow-up paper \cite{Stenflo2006} 128 citations), so it is fair to quote from it to describe the concept of QDP: ``...when a dusty  plasma is cooled to an extremely low temperature....ultracold dusty plasma behaves like a Fermi gas...'' The authors presented a quantum hydrodynamic treatment of the electron, ion and dust component of the QDP and
studied the dispersion relation of the dust acoustic wave. 
They found that the results are significantly affected by quantum  corrections and concluded that their results ``can be helpful for diagnostics of charged dust impurities in microelectronics''. 

  The work \cite{Ali2005} has been followed by a large number of papers on QDP that were 
dedicated to important plasma physics phenomena, such as the Jeans instability 
\cite{Rozina, Jamil2017, Stenflo2006, Sharma,  Jamil, Jamil2}, different types 
of dust ion acoustic waves   \cite{Ali_NJP, Karmakar, Rubab, Luo, El-Labany2010, 
Wang, Hossain, Hossain2, Sharma_EPL, Han, El-Taibany, Kumar, Kohli, Masood, 
Misra, Emadi, Chatterjee,Mamun,Rouhani, El-Labany, Ghorui, Bagchi, Abdelsalam}, 
the dust-lower-hybrid instability \cite{Rozina2, Jamil3}, new low-frequency 
oscillations \cite{Stenflo2006_EPl}, plasma waves and instabilities under 
a gravitational field \cite{Ayub, Jain}, screening effects \cite{Zeba}, and 
nonlinear ion acoustic waves \cite{Dubinov}.
These and related studies demonstrated that the inclusion of charged dust 
particles leads to a broad set of waves and oscillations that is much richer 
than in ``ordinary'' electron-ion quantum plasmas (see e.g. 
Refs.~\cite{Stenflo2006_EPl, Abdelsalam, Kumar}). In addition, it was shown that
 the presence of the charged dust particles significantly affects the 
parameters of the  acoustic waves and ionic solitons \cite{Dubinov} 
 in quantum plasmas due to modification of the charge balance between ions and electrons.
 On the other hand, the results that are well-known from  classical (dusty) 
plasmas get modified due to incorporation of quantum corrections. 
 For example,  quantum effects have been reported to stabilize the Jeans 
instability \cite{Stenflo2006, Rozina, Jain}, and dust acoustic waves are modified due to quantum effects \cite{Ali_NJP, 
El-Taibany, Hossain, Hossain2}.
  \begin{table}[h]
      \centering
      \begin{tabular}{|c|c|c|c|}\hline
          Reference & Plasma density & Temperature & Dust parameters \\
           & $n_e[cm^{-3}]$ & $T[K]$ & $A: a_D[\mu m]$\\
           \hline\hline
           \cite{Ali2005}& ? &? &?\\
           \cite{Rozina2}& $10^{25}-10^{27}$ &  $10^5$& C: ?\\
           \cite{Zeba}& $10^{24}$&$300$ &?:?\\
           \cite{Jamil3}& $10^{24}$& ? &?:$10^{-2}$ \\
           \cite{Jain}& $10^{27}$& $10^8$ &?:? \\
           \cite{El-Taibany}& $5\times10^{23}$& $100$ &?:? \\
             \cite{Ali_NJP}& $5.9\times10^{22}$& $6.4\times10^4$ &?:? \\
             \cite{Rubab}& $10^{23}$& $10^4-10^5$ &?:? \\
           \cite{Wang}& $5\times10^{29}$& 100 &?:? \\\hline
      \end{tabular}
      \caption{Parameters of quantum dusty plasmas that were used in the 
references listed in the left column which are typical examples. ``A'' 
denotes the chemical element and $a_D$ the radius of the dust particles. 
Missing information is indicated by question marks.}
      \label{tab:qdp-parameters}
  \end{table}
  
  In these papers quantum dusty plasmas with a broad range of parameters were 
considered, a few characteristic examples are listed in 
table~\ref{tab:qdp-parameters}. From this table one can see that the temperatures that were considered range
 from cryogenic ($T\sim 100 ~{\rm K}$) to extremely hot 
($T=10^{8}~{\rm K}$), whereas the plasma density is typically assumed to be very 
high, $n\gtrsim 10^{23}~{\rm cm}^{-3}$. While this is sufficient to characterize 
the electron and ion component of the QDP (assuming a fully ionized 
isothermal plasma containing a single ion component), however, the parameters 
describing the dust particles such as their material, geometry and size, are 
often omitted which is indicated by the question marks in the table. 

The missing information on the dust particle propeties, unfortunately, makes it 
difficult to reproduce the results of the theoretical analysis or, at least,
 verify their physical relevance. This is critical since QDP 
studies have been motivated by their authors by claiming importance of 
dust particles for nearly all
quantum plasmas, including such diverse objects as white dwarf stars, 
  the outer envelope of neutron stars, as well as metals and micro- and 
nano-electromechanical devices.
Thus, there should be ample opportunities to verify the theoretical predictions 
on QDP experimentally -- at least in laboratory quantum plasmas. However, to 
date -- to the best of our knowledge -- not a single experimental 
observation of the existence of dust particles in a quantum plasma has been 
reported.



This is very surprising considering the strong effect dust particles 
are known to have on the properties of classical plasmas and the achieved good agreement of experiments with the theoretical predictions in that field, 
e.g. \cite{loewen-morfill,bonitz_prl_06,hanno_prl_12} which indicates an overall good 
understanding of these systems.
So why are there no experimental results on QDP?
A possible explanation 
is the lack of precise predictions of relevant 
situations, in theoretical papers. 
%
%
Many of these works performed a largely formal analysis of the properties of 
(typically dimensionless) equations of motion for oscillations and waves. They 
did not perform a critical analysis of how the values of the
dimensionless parameters for which an effect is found in the 
simulations translate into the parameters of a ``real'' physical system and what 
specific system could be a candidate for an experimental verification of the 
made predictions about QDP effects. 

Note that, in contrast to classical dusty plasmas, that 
can often be understood -- at least qualitatively -- from the behavior of 
electron--ion plasmas by replacing (one) ion component by a very heavy strongly 
negative ``ion'' and rescaling masses and charge numbers, such a procedure 
is not guaranteed to work for dense quantum plasmas. And, as we will show, this 
procedure, in fact, breaks down for ``quantum dusty plasmas''.
%
In this paper we present a critical review of the parameters of quantum dusty 
plasmas and analyze the involved assumptions. The analysis reveals that dust 
particles cannot co-exist with quantum electrons or ions: ``quantum dusty 
plasmas'' do not exist in reality.
  
  This paper is organized as follows: in Sec.~\ref{s:parameters} we introduce 
the plasma parameters, both dimensional and dimensionless, that were considered 
in previous works on quantum dusty plasmas.
  In Sec.~\ref{s:QDP_examples} we discuss often used examples of QDP. After 
that,  in Sec.~\ref{s:3}, the stability of a dust particle that is immersed 
into a quantum plasma is examined whereas, in 
Sec.~\ref{s:dust_formation}, we inquire whether a dust particle can form and 
grow in a quantum plasma environment. Finally, the credibility of the used QDP 
parameters is discussed in Sec.~\ref{s:4}.
  \section{Quantum dusty plasmas. Relevant DIMENSIONLESS PARAMETERS}\label{s:parameters}
Let us start by giving a definition of a quantum dusty plasma. Following the concept of Ref.~\cite{Ali2005} we formulate three requirements:
\begin{description}
\item[I.] The plasma should be quantum degenerate. Compared to Ref.~\cite{Ali2005} we relax this condition by requiring only that the electrons are quantum degenerate.
\item[II.] The quantum plasma should contain stable dust particles -- particles of micrometer (or at least nanometer) diameter. 
\item[III.] Dust particles are clearly distinct from atoms, ions and molecules by their (significantly bigger) size and charge. To be specific we will require that a dust particle contains at least $N_A=10^3$ atoms. However, this number is not critical for our results obtained in Sec.~\ref{s:dust_formation} which are also valid for $N_A=1$.
\end{description}
We now recall the dimensionless 
parameters that  characterize the relevance of quantum effects and also the 
strength of correlations (Coulomb interaction) in a two-component electron-ion 
plasma, e.g. \cite{bonitz_qkt}:
\begin{enumerate}
    \item the electron degeneracy parameters $\theta_e = k_B T_e /E_F $ and 
$\chi_e=n_e\Lambda_e^3$, where $ \Lambda_e=h/\sqrt{2\pi m_e k_B T_e}$, is 
the thermal DeBroglie wave length and $E_F$ is the Fermi energy of electrons;
    \item the ion degeneracy parameter $\chi_i=n_i\Lambda_i^3$, where $\Lambda_i=h/\sqrt{2\pi m_i k_B T_i}$. Obviously the ion degeneracy parameter is a factor $(m_eT_e/m_iT_i)^{3/2}$ smaller than the one of the electrons;
    \item the quantum coupling parameter (Brueckner parameter) of electrons, $r_s = a_e/a_B$, where $a_e=(4/3\pi n_e)^{-1/3}$, and $a_B$ is the first Bohr radius;
    \item the classical coupling parameter of ions $\Gamma_i= Q_i^2/(a_ik_BT_i)$, where $Q_i$ is the ion charge, and $a_i$ is the mean inter-ionic distance.
    \item the degree of ionization of the plasma: the ratio of free electrons to the total (free plus bound) electron number, $\alpha^{ion}=n_e/n_{tot}$, determines how relevant plasma properties are.
\end{enumerate}
The presence of dust particles introduces additional parameters to the system. Guided by the experience with classical dusty plasmas, we consider: 
\begin{description}
\item[6.] the dust particle parameters: the particle radius and charge number 
$a_D, Z_D$, the atomic species ``A'' and mass density $\rho_D$, and the binding 
energy $E_D$ of atoms to the dust particle;
\item[7.] the Havnes parameter $P=|Z_d|n_D/n_e$, 
\item[8.] the dust particle degeneracy parameter $\chi_d=n_d\Lambda_d^3$, which is a factor $(m_eT_e/m_dT_d)^{3/2}$ smaller than the one of the electrons [cf. 2. above]. For an isothermal plasma and carbon dust of minimal size (cf. III.) we have $\chi_d/\chi_e \approx 10^{-9}$;
\item[9.]  the dimensionless dust particle surface potential, 
$\phi_s=-Z_de^{2}/a_D k_BT_e$, which, for a quantum plasma, can be replaced by $\phi_s=-Z_de^{2}/a_D E_{Fe}$.
\end{description}


The simplest approximation to evaluate the dust particle surface potential, 
$\phi_s$, is the orbital motion limited theory (OML) \cite{bonitz_rpp10}. 
Equating the flux of the electrons and the flux of the ions, $\phi_s$ can be evaluated from the solution of the following equation:
\begin{equation}\label{eq:1}
 \sqrt{\frac{m_i}{m_e}\frac{T_e}{T_i}}\exp (-\phi_s)= \left(1+\phi_s~\frac{T_e}{T_i}\right),
\end{equation}
where electron release from the dust particle surface due to radiation is neglected.

In the special case of an isothermal plasma with $T_e=T_i$ -- which applies to two of the QDP examples, namely the atmospheres of  neutron stars and white dwarfs [cf. Sec.~\ref{s:QDP_examples}] --  $\phi_s$ obtained from Eq. (\ref{eq:1}) depends only on the mass ratio. Then, for the example of a hydrogen (helium) plasma, $\phi_s\simeq 2.5$ ($\phi_s\simeq 3$). In general, for any  atomic mass, $\phi_s\lesssim 5$.
Note that for the  non-isothermal case, the value of  $\phi_s$  weakly depends on $T_e/T_i$. For instance, at $T_e/T_i=100$, $\phi_s$ is approximately two times smaller  compared to the isothermal case. Further, 
the effect of collisions is to decrease the dust particle surface potential \cite{Lampe1, Lampe2}.
Also, $\phi_s$ is restricted from above by the tensile strength of the dust material. 
Fracturing will occur if the electrostatic stress, inside of a dust particle due to charging, $S\sim(\phi_s k_BT_e/ea_D)^{2}$, exceeds the tensile strength \cite{Draine}. 
Additionally, the grain surface potential is limited by the electron (ion) field emission \cite{Draine}.
In general $\phi_s$ may depend on many additional factors including the  plasma particle temperatures and densities, dust particle material, size and shape,
external fields (electrostatic, magnetic, radiation).  Nevertheless, the majority of studies on  dust particle charging in different plasma environments clearly show that the dimensionless dust particle surface potential, $|\phi_s|$, is on the order of $1$.
Specifically, this is true for dust charging by external radiation \cite{Sickafoose}, which is important for consideration of dust in astrophysical contexts.

Even though the above results have been applied to QDP as well, one has to point out that, in the case of quantum plasmas, i.e.  $\chi_e>1$  (or $\theta_e<1$), this has to be questioned.
 In particular, one has to use Fermi-Dirac 
statistics for the description of electrons instead of Boltzmann statistics, 
and Eq.~(\ref{eq:1}) has to be modified accordingly. 
In the Appendix we show 
that $|\phi_s|$ is, indeed, on the order of $1$ even in the case of degenerate electrons in a quantum plasma.  


  \section{Commonly cited examples of QDP }\label{s:QDP_examples}
\begin{figure}[t]
\includegraphics[width=0.48\textwidth]{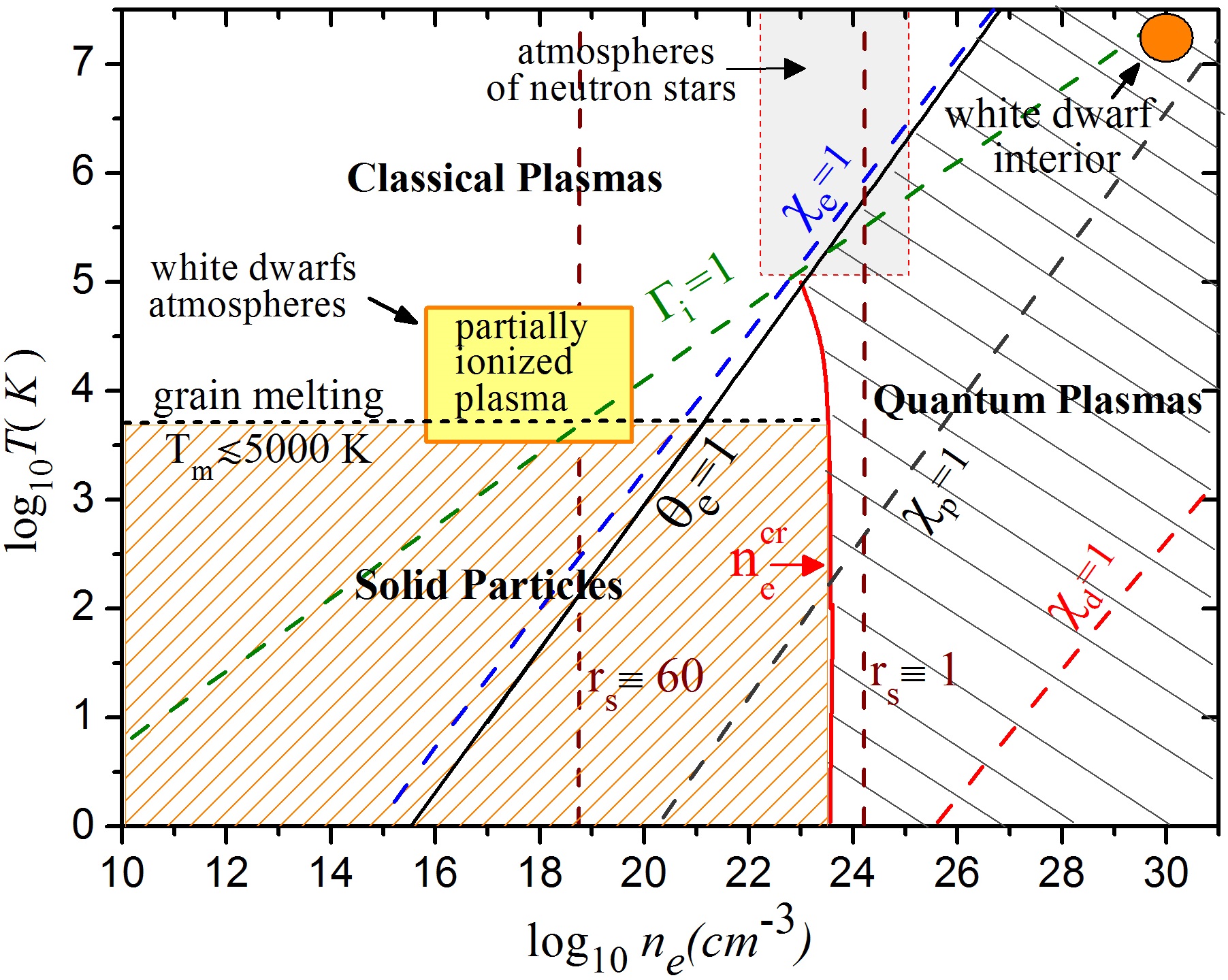}
\caption{Density-temperature plain with examples of plasmas and characteristic plasma parameters. \textit{Electronic quantum effects} are observed to the right of the line  $\chi_{\rm e} = 1$. Ion degeneracy sets in to the right of the line $\chi_{\rm p} = 1$ (for hydrogen). \textit{Dust particles can only exist} inside the orange shaded area: to the left of the line $n_e^{cr}$, Eq.~(\ref{eq:min_binding_en}) [shown for carbon], and below the grain melting line.
\textit{Dust quantum effects} for an ensemble of  particles consisting of $10^3$ carbon atoms each,  sets in below the line, $\chi_{\rm d} = 1$ and  to the left of the line $r_s=60$, i.e. for $T\lesssim 10^{-4}$K, see Sec. \ref{s:4}~\textbf{B}). For  $\chi_{\rm d} = 1$, the horizontal axis should be  understood as the density of the dust particles.}
\label{fig:1}
\end{figure}

Three examples that have been commonly used in the context of quantum dusty 
plasmas are white dwarf stars,  neutron stars,  and  micro- and nano-devices 
that are ``contaminated'' by dust particles.
  First, we consider the plasma  parameters in these objects to identify whether they are relevant to quantum plasmas.
   Temperatures and densities relevant to white dwarfs and neutron stars are 
presented in Fig. \ref{fig:1}, where we also show the characteristic lines 
$\Gamma=1$, $\chi_{\rm e(p)}=1$, $\theta_e=1$ since they separate  
different plasma regimes. Quantum effects (of electrons) are of importance to 
the right of the line $\chi_e =  1$ ($\theta_e =1$). 
 For the calculation of the ion degeneracy parameter 
the mass of the proton is used.  Note that these parameters depend only on the number of free electrons (ions). For vanishing ionization degree, $\alpha^{ion}$, the system is neutral. In an isothermal plasma,  ionization is negligible for densities and temperatures below approximately
\begin{align}
n^{ion} &\sim 10^{23} {\rm cm}^{-3}\,,    
\label{eq:nion}
\\
T^{ion} &\approx 5000 K\,.
\label{eq:tion}
\end{align}
 Outside this area the system will behave as a gas or liquid. Of course there is always a residual number of free electrons which, however, is so small that it will not exhibit quantum properties. The above estimates (\ref{eq:nion}, \ref{eq:tion}) will be derived in Sec.~\ref{s:dust_formation}.

Additionally, in Fig. \ref{fig:1} the 
horizontal line ``grain melting'' gives an estimate from above for the melting
temperature of known materials, above which no ``dust particle'' can exist. 
 Further, the line $r_s=1$ is important as close to this density 
the Mott transition (pressure ionization) occurs (see 
Sec.\ref{s:dust_formation}).

  \begin{enumerate}
   \item[\textbf{A.}] One of the motivations for studying quantum dusty plasmas was the observation of certain white dwarf atmospheres that are ``contaminated'' by dust particles \cite{Jura}.
   As observations indicate, the presence of metals within cool white dwarf 
atmospheres is due to external sources (circumstellar and interstellar).
 Dust particles can sustain in the photosphere of white dwarfs through radiative 
levitation if $T_{\rm eff}>20~000~K (30~000~K)$, for hydrogen (helium) atmospheres \cite{Chayer}, where the 
effective temperature $T_{\rm eff}$ is determined by Stefan's law 
\cite{Koester}.
 At lower temperatures, white dwarfs develop convection zones which enhance the gravitational settling of the heavy elements. 
 One of the explanations of the atmospheric pollutions is the accretion from a
dusty disk which is created by the destruction of a minor 
  body within the tidal radius of the star \cite{Debes, Jura}.   

It is important, however, to recall that the plasma parameters of the atmosphere
and the interior of white dwarfs  are strikingly different.
  The atmosphere of a cool white dwarf is a dense gas or liquid at $\sim 
10^{3}-10^{4}~{\rm K}$ 
  containing non-degenerate classical electrons with a number density 
$n_e\lesssim 10^{19}~{\rm cm^{-3}}$ \cite{Kowalski} (with the atmosphere mass 
density $\rho\lesssim3~{\rm g/cm^{3}}$). In contrast, the white dwarf 
interior contains degenerate dense electrons with $n_e> 10^{27}~{\rm cm^{-3}}$. 
   This difference is due to the  separation of heavy  elements (e.g. carbon and oxygen) from light elements (e.g. hydrogen and helium) by a strong gravitational field.

   In Fig. \ref{fig:1}, the parameters of the white dwarf's atmosphere and interior are plotted on the density-temperature plain.
   It is clearly seen that the white dwarf's atmosphere is outside of the quantum regime, whereas  the white dwarf's interior is in a quantum plasma state.
   Therefore, we exclude the  white dwarf's atmosphere from our analysis of 
quantum plasmas.
   
\item[\textbf{B.}] The atmosphere (outer envelope) of a neutron star has an
electron density in the range $10^{22}<n_e<10^{25}~{\rm cm}^{-3}$ (the mass 
density is $\rho\lesssim 10~{\rm g/cm^{-3}}$) and temperature, $T\gtrsim 10^{5}~{\rm 
K}$ \cite{Haensel}.  
  The atmosphere of a cold neutron star has a thickness of about $\sim 1~{\rm 
cm}$ and can be as thin as few millimeters when the surface temperature is 
of the order  of $\sim 10^5~{\rm K}$. 
  Even cooler neutron stars have a solid surface without an atmosphere.  Due to the high density, the atmosphere of a neutron star can be a degenerate 
or partially degenerate plasma.
  This is illustrated in Fig.~\ref{fig:1}. The parameters of the interior of neutron stars are not relevant to the current analysis.   It should be noted that no sign of dust particles has been detected in 
the atmosphere of  neutron stars.
 
\item[\textbf{C.}] 
``Contaminated'' micro- and nano-devices have been put forward as an example of 
quantum dusty plasmas in the original paper by Ali and 
Shukla \cite{Ali2005} in analogy to classical dusty plasmas.
The latter are generated in radio frequency  discharges and have been widely 
 studied  since the early 1990s. 
Initially, the interest in  dusty plasmas originated in gas discharges, in 
particular by the contamination of industrial plasma 
reactors by dust particles. 
Later, this interest was further fueled by, e.g., the potential application of 
nanoparticles and nanostructured thin films formed in plasmas. The plasma in gas discharge experiments containing dust particles
\cite{Ramazanov1, Ramazanov2, bonitz_rpp10} 
 is non-degenerate with electrons having temperatures of a few electron volts 
and densities on the order of $\sim10^{10} {\rm cm^{-3}}$.
 Ions and neutrals are, as a rule, at room temperature.  Even in the gas 
discharge at cryogenic conditions \cite{Minami, Kojima, Antipov, Kubota}, 
 with heavy particle temperatures $\sim 10~{\rm K}$,
 ions and electrons remain too hot for manifestation of quantum effects \cite{Maiorov, Polyakov}. 
The dust particles that form inside a gas discharge plasma eventually fall down on the surface of a plate (which is usually metallic in industrial plasma reactors) \cite{Boufendi2, Boufendi3}; 
thereby ``contaminating'' the micro- or nano-device.

The authors of ref.~\cite{Ali2005} suggested that further cooling of a dusty plasma will give rise to a quantum dusty plasma where the dust particles behave quantum mechanically. We will show below that this is impossible.

Of course, one can inquire about the possibility of quantum dusty plasmas in metallic devices because there the electrons typically behave as a quantum degenerate moderately correlated Fermi gas~\cite{dornheim_physrep18}. However, this gas resides inside an ionic lattice with a spacing of a few angstroem. It is impossible to ``embed'' a dust particle in this electron gas without destroying or strongly modifying the ionic lattice (e.g. by creating  defects). The resulting system is not a ``dusty plasma'' but an electron gas in a (possibly distorted) lattice, and therefore,
these examples are not included in Fig.~\ref{fig:1}.
 \end{enumerate}
From this discussion and Fig.~\ref{fig:1} 
it is clear that a quantum dusty plasmas cannot exist in the condensed matter phase but only in the gas phase. Furthermore, as discussed above, cf. Eq.~(\ref{eq:nion}) and Fig.~\ref{fig:1}, 
in the gas phase the electron density has to exceed $n_e\gtrsim 10^{23}~{\rm cm^{-3}}$ for the electrons to be quantum degenerate.


Now we proceed further and show that a dust particle is not able to survive in the plasma with an electron density $n_e\gtrsim 10^{23}~{\rm cm^{-3}}$. 
 
\section{Dust grain destruction mechanism under dense quantum plasma conditions} \label{s:3}
Let us start from an analysis of the conditions 
that must be fulfilled for a dust particle to be stable in a quantum plasma. 
This is equivalent to finding conditions where destruction mechanisms of dust particles are not active (or slow enough).
The most important mechanisms which lead to the  destruction of a dust particle due to the fluxes of ions and electrons to the surface of the grain 
are:
  \begin{enumerate}
\item[(i)]  melting and evaporation of the dust material due to heating, 
\item[(ii)]  sputtering of the dust material,
\item[(iii)]  field emission of ions from the dust particle.
\end{enumerate}
We do not consider dust material sublimation, which is too slow for our consideration, even though it is important on astrophysical time scales \cite{Kama}. 

The mechanism (i) 
takes place when the cooling of a dust particle  
by neutral atoms, radiation, thermal emission of electrons and ions is unable
to stabilize the dust material temperature. 
Mechanisms, (ii) and (iii), are important at high energies of ions colliding with the dust particle surface \cite{Draine, Vignitchouk}.
In the context of this paper, ``high temperature'' means that the ion (atom) temperature exceeds the melting temperature of the dust particle, $T_i>T_m$.
At high temperatures and low plasma density ($n_e\sim 10^{12}-10^{14}~{\rm cm^{-3}}$), 
the investigation of a micron size dust particle evolution in the plasma
 of tokamak fusion devices showed that the life time of a dust particle, $\tau$, ranges from $\tau \sim 10^{-4}~{\rm s}$ to $\tau \sim 0.1~{\rm s}$, depending on the initial size of the grain \cite{Vignitchouk, Bastykova}.
 Comparing this to quantum plasmas which have similar temperatures but much higher densities, $\gtrsim 10^{23}~{\rm cm^{-3}}$, 
 the dust particle life time would be much shorter, 
 because the energy flux to the surface of a dust particle, which is proportional to $n_e$, is larger billions times.  
 
 At  lower temperatures and densities in the range of  $10^{23}~{\rm cm^{-3}}$ and $10^{24}~{\rm cm^{-3}}$,  calculations  based on the model of Ref.~\cite{Vignitchouk} that takes into account all important heating and cooling mechanisms as well as sputtering, yields $\tau \ll 1 ~{\rm s}$.
 For example, at $T=10~ {\rm K}$, 
 for a micron sized dust particle [relevant dust materials are tungsten, graphite, and silicates], $\tau < 1$ ns, whereas at $1000~{\rm K}$, $\tau \lesssim 0.1~{\rm ns}$.
 At still higher temperature and fixed density, $\tau$ is obviously even shorter, as the dust destruction become more efficient with increase of
 temperature. 
 Therefore, we conclude that the atmosphere of a neutron star is not a canditate for quantum dusty plasmas.

 Before proceeding further let us discuss in more detail why a dust particle cannot survive in plasmas with densities, $n_e>10^{23}~{\rm cm^{-3}}$.
Recall that the dust particle surface temperature is stabilized most effectively via cooling by neutral atoms,  and by the radiative energy loss \cite{Khrapak_POP, Delzanno_POP}.
 At these densities which are close to the Mott transition or beyond the degree of ionization approaches one, and  
 the cooling by neutral atoms is not relevant. 
 For the dust surface temperature to be stabilized,
 the heat flux to the grain surface $\Gamma_H$  due to 
 energy deposition of collected ions and electrons and their recombination should equal the energy loss flux due to radiation, $\Gamma_R$.
 The former is approximated as $\Gamma_H\thickapprox J_0k_BT_e(2+\phi_s+I/k_BT_e)$, where $I$ is the ionization energy, and the plasma flux is approximated in OML, $J_0\simeq \sqrt{8\pi}a_D^2n_ev_{T_e}\exp(-\phi_s)$ \cite{Khrapak_POP}.
 The radiation flux is treated as black body radiation,
  $\Gamma_R\thickapprox 4\pi a_D^{2}\sigma \left(T_s^4-\zeta T_e^4\right)$, where 
 $\sigma$ is the Stefan-Boltzmann constant and
  $\zeta>1$ is a correction due to a positive shift of the radiation frequency in a plasma, $\omega^2=c^2k^2+\omega_p^2$ \cite{Ichimaru}. 
 Assuming stability of the surface temperature, we find:
  \begin{equation}\label{eq:2}
 T_s^{4}\gtrsim n_e T_e^{3/2}\frac{k_B^{3/2}}{\sqrt{2\pi m_e}\sigma}e^{-\phi_s}(2+\phi_s+I/T_e)+\zeta T_e^4,
\end{equation}
where $\left(\frac{k_B^{3/2}}{\sqrt{2\pi m_e}\sigma}\right)^{1/4}\simeq 2.5\times 10^{-2}$ (in {\rm CGS} units).  

Using Eq.~(\ref{eq:2}) with $\phi_s\sim1$ and $k_BT_e\gtrsim I$, 
we find that radiation is unable to prevent dust particle melting as $T_s>\left(10^{16}\times T_e^{3/2}+ T_e^4\right)^{1/4}$, at  $n_e\gtrsim10^{23}~{\rm cm^{-3}}$.
Indeed, even if we minimize thermal effects by assuming an unrealistically low plasma temperature, $T_e=1~{\rm K}$, we find that $T_s>10^{4}\times T_e^{3/8}\simeq 10^4 ~{\rm K}$ and, at the considered high densities, the dust particle surface temperature is well above the melting temperature of all known materials. At a more realistic plasma temperature, $T\sim 10^4~{\rm K}$, we find that $T_s>10^5~{\rm K}$.
 Therefore, at these high densities,
melting of the dust particle is unavoidable. Melting, in turn,  facilitates evaporation of atoms, sputtering, and field emission from the surface and rapid destruction of the dust particle.
Indeed, the heat flux per dust particle atom, $\Gamma_H/N$, exceeds the binding energy of the surface atoms, $E_D$, which is on the order of $10~{\rm eV}$,
and one easily finds that a micron sized dust particle of the considered materials will loose every single atom within less than $1~{\rm ns}$.

  To summarize, we conclude that dust particles cannot survive at $n_e\gtrsim10^{23}~{\rm cm^{-3}}$, 
  particularly in the atmospheres of neutron stars and interiors of white dwarfs.

 \section{Dust grain formation in a quantum plasma}\label{s:dust_formation}
 We may turn the question around and ask under what conditions a dust particle can form in a quantum plasma.
 The first necessary step is, obviously, that a neutral atom can form from the electron-ion plasma.
 Only after this has happened and the density of atoms is sufficiently high, aggregation of atoms into dimers and larger complexes and, eventually into a ``dust particle'', can occur.
  \begin{figure}[t!]
     \centering
\includegraphics[width=0.45\textwidth]{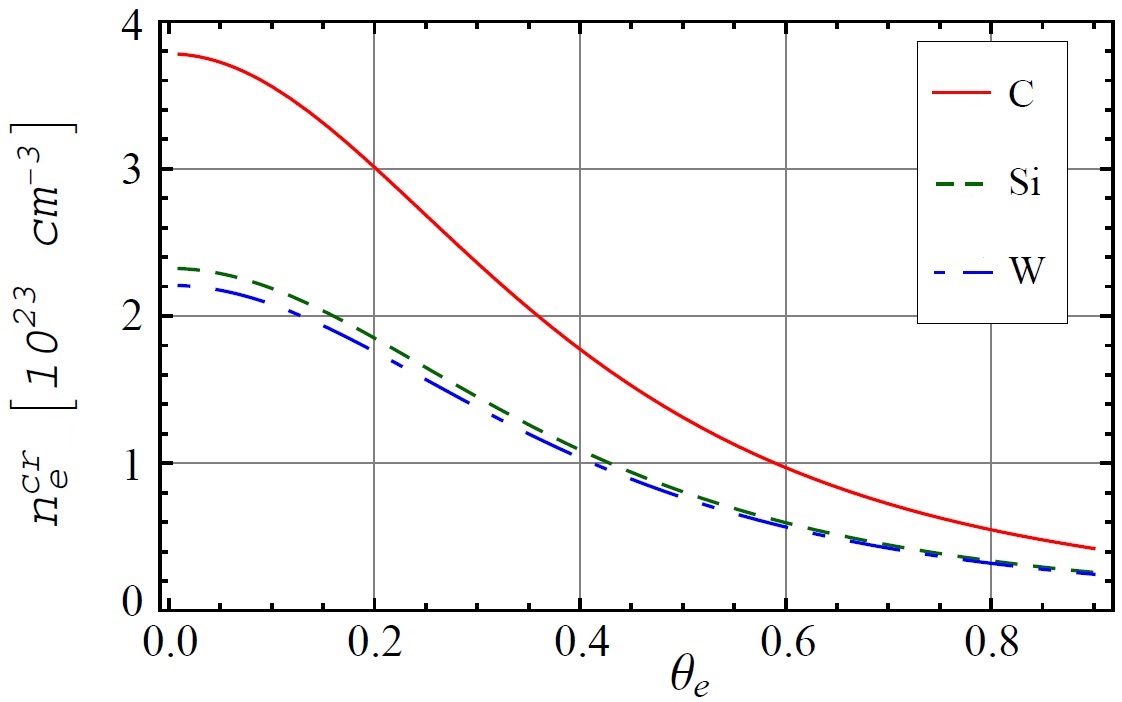}
     \caption{Critical values of the  electron density calculated according to Eq.~(\ref{ncr}) vs. electronic degeneracy parameter, 
     for carbon (C, $\alpha=0.83$), silicon (Si, $\alpha=0.6$), and tungsten (W, $\alpha=0.58$). 
     For $0.1<\theta_e<1$, the temperatures corresponding to the calculated critical densities are in the range 
      $2\times 10^4~{\rm K}<T<4.2\times 10^{4}~{\rm K}$, for C,  
      $1.5\times 10^4~{\rm K}<T<3\times 10^{4}~{\rm K}$, for Si,
      and $1.4\times 10^4~{\rm K}<T<2.9\times 10^{4}~{\rm K}$, for W.}
     \label{fig:binding_en}
 \end{figure}
 For the formation of the seed neutral atom it is necessary that the binding energy of the dust atoms exceeds the kinetic energy of the electron to be captured by the seed ion, $E^{(1)}_D > E^e_{\rm kin}$. Assuming thermal equilibrium, in a classical system this would amount to requiring $E^{(1)}_D > \frac{3}{2}k_BT_e$. For the example of hydrogen, this temperature is of the order of $10^5$K. The next step -- the formation of a dimer (e.g. $H_2$ molecule) -- requires already a much lower threshold for the electron temperature on the order of $30,000$~K due to the lower molecule binding energy. With increasing cluster size the binding energy of the next atom converges to a value on the order of $E_D^{(\infty)} \sim 1...10$~eV, corresponding to $T_e\sim 10,000...100,000$~K, for most materials. This value for the electron temperature exceeds the melting temperature of the dust particles, $T_m \lesssim 5,000$~K which, therefore, sets the upper threshold for the existence of dust particles, in a classical dusty plasma, cf. Fig.~\ref{fig:1}.
 
Let us now turn to a quantum plasma. In the extreme case of strong electron degeneracy, $T_e=0$, and the mean kinetic energy is given by the Fermi energy. So the above condition for bound state formation changes to $E_D^{(1)} > \frac{3}{5}E_{Fe}$. At finite temperature the expression can be corrected via a Sommerfeld expansion. Retaining terms of lowest order in $\theta_e$ we obtain:
 \begin{align}
     E_D^{(1)} > \frac{3}{5}E_{Fe} \left\{
     1 + \frac{5\pi^2}{12}\theta_e^2
     \right\}\,.
     \label{eq:min_binding_en}
 \end{align}
 For example, for a hydrogen atom, $E_D^{(1)}=E_H=13.6$~eV, and it is straightforward to compute the maximal density a hydrogen atom can withstand in a low-temperature quantum plasma ($\theta_e=0$). From Eq.~(\ref{eq:min_binding_en}) one readily finds the critical value of the Brueckner parameter, $r_s^{cr}(T=0) \approx 1.5$ which corresponds to an electron density $n_{cr}(T=0)\simeq 5\times 10^{23}~{\rm cm}^{-3}$  which is in reasonable agreement with quantum Monte Carlo results \cite{BonitzPRL}.

At finite temperature we obtain from Eq.~(\ref{eq:min_binding_en}), 
\begin{eqnarray}\label{ncr}
r_s^{cr}(\theta_e) &\approx & \frac{1.5}{\alpha^{1/2}}\sqrt{1+\frac{5\pi^2}{12}\theta_e^2}, 
\\
n_{e}^{cr}(\theta_e) &\approx &   \left(\frac{\alpha}{\sqrt{1+\frac{5\pi^2}{12}\theta_e^2}}\right)^{3/2} 5\times10^{23}~\left[{\rm cm}^{-3}\right],
\end{eqnarray}
where $\alpha$ is the ratio of the binding energy of the considered ion (atom) to $E_H$. 

 For heavier elements with the nuclear charge $Z$ the binding energy of the first electron to the nucleus equals $Z^2 \times 13.6$~eV, giving rise to a $(Z-1)$-fold charged ion. However, with each subsequent electron the binding energy rapidly decreases until it reaches values on the order of the hydrogen binding energy, for the last electron.
 For example, for the important case of carbon, the binding energy of the first electron (formation of the C$^{5+}$-ion) is approximately $490$~eV whereas for the neutral atom it is $11.26$~eV. As discussed above, only after a neutral atom has been formed, eventually, particle growth can set in.
 The critical density given by Eq.~(\ref{ncr}) is plotted in Fig.~\ref{fig:binding_en} 
 for three different types of atoms.
 From this figure one can see that, at $\theta_e<1$ and $n\gtrsim10^{23}~{\rm cm}^{-3}$, the neutral atoms C and Si, which are often considered to be building blocks of the dust particles, cannot form. The same conclusion is also valid for W, which has the highest melting point of all elements. The line 
 (\ref{ncr}) is also plotted in Fig.~\ref{fig:1}, for the example of carbon particles.
 
Finally, the above analysis  of bound state formation can also be applied to analyze the degree of ionization in the plasma. Indeed, if we consider the binding energy $E^{(P,1)}_D$ of the plasma gas atoms. 
 In fact, in a classical plasma, the ionization probability is proportional to $\alpha^{ion}(T)\sim \exp(-E_D^{(1)}/k_BT)$ \cite{schlanges_cpp_95}. Therefore, for relevant plasmas, ionization is negligible for temperatures below approximately $T^{ion}=5000$~K, in agreement with Eq.~(\ref{eq:tion}). On the other hand, in a quantum plasma we can use Eq.~(\ref{ncr}) as an estimate for the ionization threshold. Indeed, using again  $E^{(1)}_D$ for the ionization energy of the plasma atoms, a quantum plasma can only exist to the right of the corresponding line, Eq.~(\ref{ncr}), for the plasma atoms which justifies the estimate given in Eq.~(\ref{eq:nion}) above. The corresponding line for the dust particles has to use, instead, the binding energy for larger particles, $E_D^{(\infty)}$, which is, in most cases, lower than $E_D^{(P,1)}$ for relevant plasmas. This again confirms that the existence of dust particles is incompatible with the existence of a quantum plasma, as indicated in Fig.~\ref{fig:1}. 
 
  \section{Test of quantum  dusty plasma parameters used in the literature} \label{s:4}
    To complete the picture of current QDP studies,
    we critically review some examples of dimensionless parameter values and assumptions that have been used in the literature.
    This will show that, often, contradictory and even unphysical conditions have been considered.
  
    \begin{enumerate}
  \item[\textbf{A.}]  Many papers, e.g.  \cite{Rubab,Rozina2,Jain,AliEPJ,Ali_NJP} consider high plasma temperatures way above the stability limits of any material as we demonstrated in Sec.~\ref{s:dust_formation}. Some examples have been listed in table~\ref{tab:qdp-parameters}. 
  
  \item[\textbf{B.}]  Many papers consider very strongly degenerate electrons with $\theta_e\ll1$, e.g. 
  \cite{Ali2005, Rozina, Jamil2017, Stenflo2006, Sharma, Jamil, Jamil2, Ali_NJP, Karmakar, Rubab, Luo, El-Labany2010, Wang, Hossain, Hossain2, Sharma_EPL, Han, El-Taibany,Kumar, Kohli, Masood, Misra,   Emadi, Chatterjee, Mamun, Rouhani, El-Labany, Ghorui, Bagchi, Abdelsalam, Rozina2, Jamil3, Stenflo2006_EPl, Ayub, Jain, Dubinov, Zeba}. 
As we have shown above, in strongly degenerate quantum plasmas dust particles cannot survive, due to the high electron density, or low temperature, and no ``dusty'' plasma can form. 

On the other hand, some papers considered plasmas with $\theta_e\gg1 $, but this has nothing to do with a quantum plasma no matter if dust particles are present or not. For example, Ref.~\cite{AliEPJ} considered an astrophysical plasma with $n_e\sim 10^{19}~{\rm cm^{-3}}$ and $T_e=10^{5}~{\rm K}$ which is way outside the quantum plasma range, cf. Fig. \ref{fig:1}.
  \item[\textbf{C.}]
  Often dust parameters are used that are clearly incompatible. For example, Ref.~\cite{Jamil3} used the following ``typical parameters...for the interiors of the neutron stars, the magnetars, and the white dwarfs...'': a dust density of $10^{18}cm^{-3}$ and a dust radius $a_d\sim 10^{-5}$cm. One readily verifies that, at this density, the mean nearest neighbor distance of two dust particles is less than $10^{-6}$cm, i.e., more than an order of magnitude smaller than the dust radius. This is clearly impossible without destroying the dust particles.
  
  \item[\textbf{D.}] Following Ref.~\cite{Ali2005}, in many works dust particles are treated as quantum degenerate fermions with the dust ``degeneracy'' parameter $\theta_D=k_BT_D/E_F^{D}\ll1$, e.g. 
  \cite{Jamil2017,  Sharma, Jamil, Jamil2, Ali_NJP, Karmakar, Wang, Sharma_EPL, Han, El-Taibany, Emadi,  El-Labany, Rozina2, Jamil3, Zeba}, where $T_D$ is the characteristic temperature corresponding to the chaotic motion of the dust particles (which is different from their surface temperature $T_s$) and $E_F^D=\hbar^2 (3\pi^2 n_D)^{2/3}/2m_D$ is the ``Fermi energy'' of the dust component,
   and $m_D$ is the mass of a dust particle.

   To consider the most favorable condition for dust quantum effects we assume a very small dust particle size of $N_A=10^3$ atoms [cf. condition III. in Sec.\ref{s:parameters}] and estimate the temperature needed for an ensemble of dust particles to become quantum degenerate. A carbon particle of this size has a radius $a_d \sim 15$~\AA $\approx 30 ~a_B$, so the mean interparticle distance between two dust particles cannot be smaller than $60~a_B$. The maximum density such a gas of dust particles can reach is, obviously, $n^{max}=(4\pi/3)^{-1}(2 a_D)^{-3}\sim 5 \cdot 10^{18}$~cm$^{-3}$, corresponding to $r_s \approx 60$. A degeneracy parameter $\chi_D=1$ is reached for a temperature $T_d^{max}\sim 0.0001$~K, see Fig.~\ref{fig:1}. Thus, only below temperatures of one hundred microkelvin and densities below $\sim 5 \cdot 10^{18}$~cm$^{-3}$ quantum effects of extremely small dust particles can be reached if they are maximally densely packed. For more realistic dust particles of bigger size and lower density, the degeneracy temperature will be even smaller.
   
   For microelectronic applications such low temperatures are highly unrealistic. Also the close packing of dust particles with distances of $30$~\AA$\:$ are not realistic where contamination is typically due to a few dust particles. 

   Moreover, the incorrectness of the Fermi gas assumption is obvious since for dust particles of that size 
a distinction of bosonic or fermionic spin statistics makes no sense because the total spin of the particle depends on the precise number of atoms $N_A$. In an ensemble of dust particles, $N_A$ unavoidably fluctuates randomly, so this ensemble would never be strictly fermionic.

   \item[\textbf{E.}] In some papers, in addition to the assumption $\theta_e\ll1$, the parameter $\mu=\left(|Z_d|n_D\right)/\left(|Z_i|n_i\right)$ for the negatively charged dust particles was assumed to vary in the range from zero up to one
  \cite{Wang, Hossain2, Han, El-Taibany, Chatterjee, Rouhani} (see the note \cite{note_P}). 
  This assumption contradicts the condition  $\theta_e\ll1$ used in these works. 
   Indeed,  $\mu \to 1$ corresponds to vanishing electron density and thus not to a quantum plasma.
\end{enumerate}

    \section{Conclusions} \label{s:5}
 We have presented a critical assessment of the concept of quantum dusty plasmas.
 We showed that the recent high activity in the theoretical investigation of QDP is based on  unjustified assumptions and unrealistic plasma parameter combinations. This research has been motivated by entirely unrealistic examples including neutron stars, white dwarf stars or microelectronics devices.
  It was demonstrated that dust particles cannot survive or form neither in a dense quantum plasma in general, nor in the mentioned example systems, in particular. The reason is that quantum electrons produce a pressure that is so high that it unavoidably destroys any micrometer and even nanometer size particles. Therefore, dusty plasmas are restricted to the regime of classical plasmas. It is obvious that, under these circumstances, the discussion of quantum degeneracy and Fermi statistics of many dust particles is meaningless and detached from reality, cf. Fig.~\ref{fig:1}. 

  It is certainly of interest to analyze what happens to classical dusty plasmas when they are cooled to low temperatures. But, as we have shown, before this becomes a quantum plasma, the dust particles are being destroyed. Moreover a treatment of ultracold classical dusty plasmas requires a careful analysis of many important processes, including the particle bombardment with plasma particles or radiation, ionization and recombination. This may lead to important new results for the physics of dusty plasmas although this will not be a ``quantum dusty plasma''.
  
  

  \section*{Acknowledgments}
    We thank N. Bastykova (Almaty) for providing data on the dust particles' life time in plasmas.
Zh. Moldabekov thanks the funding from the German Academic Exchange Service (DAAD).
   This work has been supported by the Ministry 
of Education and Science of the Republic of Kazakhstan via the grant  BR05236730 
 ``Investigation of fundamental problems of Phys. Plasmas and plasma-like media'' (2019).  

   \section*{Appendix: Surface potential of a dust particle in a quantum plasma}
   
Here we show that, if a dust particle would exist in a quantum plasma, its dimensionless surface potential would be on the order of unity, as discussed in Sec.~\ref{s:parameters}.

  In dense quantum plasmas (see Fig.~\ref{fig:1}), the diameter of a micro- or nano-particle is much larger than characteristic plasma length scales such as the mean interelectronic distance and the screening length (Thomas-Fermi screening length). Therefore, the flux (total current) of electrons and ions through a closed spherical surface of radius $r$ around a dust particle can be computed using the drift-diffusion (extended Mermin) approximation:
  \begin{equation}\label{eq:a1}
      J_{i(e)}= 4\pi r^2\left(\pm Z_{i(e)}|e|\mu_{i(e)}n_{i(e)}E-Z_{i(e)}|e|D_{i(e)} \frac{\partial n_{i(e)}}{\partial r}\right),
  \end{equation}
  where the upper and lower sign corresponds to the flux of ions and electrons, respectively. In Eq.~(\ref{eq:a1}), $\mu_{i(e)}$ denotes the mobility,  $D_{i(e)}$ the diffusion coefficient,  $Z_i$ the charge number of an ion, and $Z_e=1$. For classical ions, Einstein's relation applies: $D_i=\mu_{i}k_B T_i/(|Z_i e|)$. For quantum electrons, the analogue of the classical  Einstein relation is more complicated, e.g. \cite{Kleinert, Marshak}. In the case of strong degeneracy, $\theta_e\ll1$, one can use $D_e\simeq (2/8)(E_F\mu_e/|e|)$ \cite{Marshak}. In a stationary state ($\partial n_{i(e)}/\partial t=0$),  the total flux is constant, i.e.,  $J_{i(e)}=const$. Assuming that all electrons and ions colliding with the dust particle surface are absorbed  (recombined), we have the boundary condition $n_{i(e)}(r=a_d)=0$. From this and, taking $E=\frac{Z_de}{r^2}$ in the vicinity of the  dust particle, we find the solution of Eq.~(\ref{eq:a1}):
  \begin{multline}\label{eq:a2}
      n_{i(e)}(r)=\pm \frac{J_{i(e)}}{4\pi Z_d e^2\mu_{i(e)}Z_{i(e)}}\times\\
      \left[1-\exp\left\{\pm \frac{Z_d|e|\mu_{i(e)}}{D_{i(e)}}\left(\frac{1}{a_d}-\frac{1}{r}\right)\right\}\right].
  \end{multline}
 Using the second set of boundary conditions, $Z_in_i(r\to \infty)=n_e(r\to \infty)=n_0$, for the total flux, we find from Eq.~(\ref{eq:a2}):
 \begin{equation}\label{eq:a3}
     J_{i(e)}=\frac{\pm 4\pi Z_d e^2\mu_{i(e)}n_0}{1-\exp\left\{\pm \frac{Z_d|e|\mu_{i(e)}}{D_{i(e)}}\left(\frac{1}{a_d}-\frac{1}{r}\right)\right\}}.
 \end{equation}
 
 The non-linear equation for the dust particle charge follows from Eq.~(\ref{eq:a3}) recalling that, in a steady state, the total current of electrons is equal to that of ions. For the quantum plasma with classical ions and degenerate electrons, $E_F\gg k_B T_{e(i)}$, we derive the charge number of the dust particle:
 \begin{equation}\label{eq:a4}
     Z_d\simeq -\frac{a_d}{|e|}\frac{D_e}{\mu_e}\,\mathrm{ln} \left(\frac{\mu_e}{\mu_i}+1\right), 
 \end{equation}
 from which  we find for $\phi_s=-Z_de^{2}/a_D E_{Fe}$ (see Sec.~\ref{s:parameters}):
 \begin{equation}
     \phi_s\simeq \frac{|e|}{E_F}\frac{D_e}{\mu_e}\,\mathrm{ln} \left(\frac{\mu_e}{\mu_i}+1\right).
 \end{equation}
 For $Z_i=1$ ($Z_i=10$), assuming $\frac{\mu_e}{\mu_i}\simeq \frac{m_e}{m_i}$, and taking $D_e\simeq (2/8)(E_F\mu_e/|e|)$ \cite{Marshak}, we obtain $\phi_s\simeq 1.9$ ($\phi_s\simeq 2.5$), in agreement with the discussion in Sec. \ref{s:parameters}. Note that, for classical electrons, $D_e=\mu_{e}k_B T_e/|e|$, and Eq.~(\ref{eq:a4}) reproduces the result for a low temperature classical plasma with $T_e\gg T_i$ \cite{Fortov_book}.
 
  




\begin{thebibliography}{10}
\bibitem{note_P}   In Ref.~\cite{Wang}, $n_i/n_e=4$ was considered (with $Z_i=1$), meaning that $P=3$ was used. In Ref.\cite{Hossain2}, for the description of the plasma composition, some other different dimensionless parameters were used instead of the Havnes parameter. Using these parameters one can deduce that the Havnes parameter in the range $0.16\lesssim P\lesssim 1.5$ was used. In Ref.~\cite{El-Taibany}, $n_i/n_e=2$ and $n_i/n_e=4$ were used, which correspond to $P=1$ and $P=3$, respectively.
\end{thebibliography}
\end{document}